\documentclass{elsart}
\usepackage{epsfig}
\def\be{\begin{equation}}
\def\ee{\end{equation}}
\def\vh{\varphi}
\newcommand{\bea}{\begin{eqnarray}}
\newcommand{\eea}{\end{eqnarray}}

\begin{document}
\begin{frontmatter}
\title{Cosmological Particle Creation and \\ 
Baryon Number Violation in a \\
Conformal Unified Theory}
\author[dub,ur]{D.~Blaschke}, \author[dub]{A.~Gusev}, 
\author[dub]{V.~Pervushin},
and \author[dub]{D.~Proskurin}
\address[dub]{Bogoliubov Laboratory for Theoretical Physics, 
Joint Institute for Nuclear Research, 141980 Dubna, Russia}
\address[ur]{Fachbereich Physik, Universit\"at Rostock, D-18051 Rostock, 
Germany}
\begin{abstract}
We consider a conformal unified theory as the basis of
conformal-invariant cosmological model where
the permanent rigid state of the universe is compatible
with the primordial element abundance and supernova data.
We show that the cosmological creation of vector Z and W bosons,
in this case, is sufficient to explain the CMB temperature (2.7 K).
The primordial
bosons violate the baryon number in the standard model as a
result of anomalous nonconservation of left-handed currents
and a nonzero squeezed vacuum expectation
value of the topological Chern-Simons functional.
\end{abstract}
\end{frontmatter}

\section{Introduction}
To explain the origin of the observable matter in the universe and
the violation of baryon number are the urgent problems
in modern cosmology. The latter is needed for the explanation of
the baryon asymmetry in the universe in accordance with the three
Sakharov conditions: CP-violation in weak interaction,
nonstationary universe, and baryon asymmetry~\cite{sakh}.

In the present paper, we try to explain the cosmological origin
of the observable matter and the violation of the baryon
number in a conformal-invariant version of the unified theory of
gauge fields and gravitation given in a space-time with the
Weyl geometry of similarity~\cite{114,plb}.
This geometry allows us to measure the ratio of two intervals and
leads to  conformal cosmology~\cite{N,039}.
In the conformal cosmology, the modern supernova data
on the accelerating universe evolution~\cite{ps} are compatible
with the dominance of the rigid state~\cite{039}. The conformal
version of the rigid state reproduces the z-history of chemical
evolution of the element abundance in standard cosmology,
since we have, in conformal cosmology, the same square root
dependence of the scale factor on the observable time in the rigid
stage. At the beginning of the universe, conformal cosmology
describes intensive creation of W and Z bosons from vacuum.
The primordial bosons violate the baryon number due to their
polarization of Dirac vacuum.

\section{Model}
We consider a conformal-invariant unified theory of gravitational,
electroweak, and strong interactions: the Standard Model where
the dimension parameter in the Higgs potential is replaced by the
dilaton field $w$ described by the Penrose-Chernikov-Tagirov action
with the negative sign. The corresponding action  takes the form
\be \label{Cw}
W=\int d^4x \sqrt{-g}\left[
{\it L}_{\Phi,w}+
{\it L}_{g}+{\it L}_{l}+
{\it L}_{l\Phi}+...\right],
\ee
where the Lagrangian of dilaton and Higgs fields is given by
\be \label{LP}
{\it L}_{\Phi,w}=\frac{|\Phi|^2-w^2}{6}R-
\partial_{\mu}w\partial^{\mu}w
+D^{-}_{\mu}\Phi(D^{\mu,-}\Phi)^*
-\lambda\left( |\Phi|^{2}- y^{2}_{h}w^{2} \right)^{2},
\ee
where $\Phi = \left(\begin{array}{rr} \Phi_{+} \\ 
\Phi_{0}\end{array}\right)$ is the Higgs field doublet,  
$ D_{\mu}^{-}\Phi=(\partial_{\mu} -
\imath g \frac{\tau_{a}}{2}A_{\mu}-
\frac{\imath}{2}g^{\prime}B_{\mu})\Phi$, and 
$|\Phi|^{2} = \Phi_{+}\Phi_{-} + \Phi_{0}\bar{\Phi}_{0}$.
The Lagrangian of the gauge fields is
\be \label{LG}
{\it L}_{g}=-\frac{1}{4}\left(\partial_{\mu}A^{a}_{\nu} -
\partial_{\nu}A^{a}_{\mu}+g\varepsilon_{abc}A^{b}_{\mu}A^{c}_{\nu}
\right)^{2}-\frac{1}{4}\left(\partial_{\mu}B_{\nu}
- \partial_{\nu} B_{\mu} \right)^{2}
\ee 
whereas that of the leptons is
\bea \label{LL}
{\it L}_{l}=\imath \bar{L} \gamma^{\mu}D^{+}_\mu L
+ \imath \bar{e}_{R} \gamma^{\mu}\left(D^F_{\mu} +
\imath g^{\prime}B_{\mu}\right)e_{R} +
\bar{\nu}_{R}\imath \gamma^{\mu}\partial_{\mu}\nu_R
\eea
with
\be \label{L}
L = \left(
 \begin{array}{rr}
 \nu_{e,L} \\
  e_L \end{array}\right);~~e_R,\nu_{e,R},
\ee
$D^F$ is the Fock derivative, $D^{+}_\mu L =(D^F_{\mu} -
\imath g \frac{\tau_{a}}{2}A_{\mu}
+ \frac{\imath}{2}g^{\prime}B_{\mu})L$.

The Lagrangian describing mass terms of leptons including that of the
neutrino (if it has one) is
\be \label{LVh}
{\it L}_{l\vh}= - y_{e}\left(e_{R}\Phi^{+}L
+ \bar{L}\Phi e_{R}\right)
- y_{\nu}\left(\bar{\nu}_{R}\Phi^{+}_{C}L
+ \bar{L}\Phi_{C}\nu_{R} \right),~~
\ee
where
\be
\Phi_{C} = \imath \tau_{2}\Phi^{+}_{C} =
\left(\begin{array}{rr}
 \bar{\Phi}_{0} ~\\
 - \Phi_{-}\end{array}\right)
 \nonumber
\ee
and $y_f$ are dimensionless parameters.
This theory is invariant with respect to conformal transformations,
and it is given in the Weyl space of similarity with the relative
standard of  measurement of intervals as the ratio
of two intervals of the Riemannian space~\cite{we}.
The space of similarity is the manifold of Riemannian spaces
connected by the conformal transformations. This relation depends
on nine metric components and allows us to use
the Lichnerowicz conformal-invariant variables~\cite{L} and
measurable conformal-invariant space-time interval
\be
\label{dse}
(ds^L)^2=g^L_{\mu\nu}dx^\mu dx^\nu,
~~~g^L_{\mu\nu}=g_{\mu\nu}|{}^{(3)}g|^{-1/3},~~~|{}^{(3)}g^L|=1~~.
\ee
The principle of equivalence of inertial and gravitational masses
is incorporated into the conformal-invariant Higgs mechanism
through the dilaton-Higgs mixing~\cite{p1}
$w^L=\phi\cosh\chi$, $\Phi_i^L=\phi~{\bf n}_i \sinh\chi$,
$|\Phi^L|^2-({w^L})^2=-\phi^2$, ${\bf n}{\bf n}^{+}=1$,
so that the action~(\ref{Cw}) takes the form
\bea \nonumber
 {\it L}= - \frac{\phi^{2}}{6}R -\partial_{\mu}\phi
\partial^{\nu}\phi +
\phi^{2}\partial_{\mu}\chi\partial^{\mu}\chi+
{\it L}_{\rm Higgs}+y_e\phi \sinh\chi \bar e e+...,
\eea
where the Higgs Lagrangian
\bea \nonumber
{\it L}_{\rm Higgs}= -\lambda\left( |\Phi^L|^{2}
- y^{2}_{h}(w^L)^{2} \right)^{2} =
 -\lambda\phi^{4}\left[\sinh^2 \chi-y^{2}_{h}
\cosh^2\chi\right]^{2}
\eea
describes the conformal-invariant Higgs effect of
the spontaneous SU(2)  symmetry breaking
 \bea \nonumber
\frac{\partial {\it L}_{\rm Higgs}}{\partial \chi}
=0~\Rightarrow \chi_1 = 0,
~~~~\sinh \chi_{2,3} = \pm \frac{y_h}{\sqrt{1 - y^{2}_h}}\sim 10^{-17}~.
\eea

The present-day value of the dilaton in the region far from heavy
masses distinguishes the scale of the Planck mass
$\phi (t_0,x)\simeq M_{ \rm  \bf Planck}\sqrt{3/(8\pi)}$.
This fact is revealed by energy-constrained perturbation
theory~\cite{plb,pp}.

\section{Method}
The lowest order of energy-constrained perturbation theory is
formed by linearization of all  equations of motion in the class
of functions with nonzero Fourier harmonics (i.e., the "local" class
of functions) in the flat conformal space-time
\be \label{cst}
ds^2_L=d\eta^2-dx_i^2,~~~d\eta = N_0(t)dt,
~~~ N_0=[g_L^{00}]^{-1/2}.
\ee
Part of these local equations are constraints that form the
projection operators. These operators remove all superfluous degrees
of freedom of massless and massive local fields. In particular,
four local constraints as the equations for $g_{\mu\nu= 0}$  remove
three longitudinal components of gravitons and all nonzero Fourier
harmonics of the dilaton. However, the local constraints could not
remove the zero Fourier component of the dilaton
$ \vh(t)  = {\int d^3x\phi^L(t,x)}/{\int d^3x}$.
The infrared interaction of the complete set of local independent
variables $\{f\}$ with this dilaton zero mode $\vh(t)$
is taken into account exactly.
The lowest  order of the considered linearized perturbation theory
is described by the Hamiltonian form of the action~(\ref{Cw}) in this
approximation
\bea\nonumber
S_0= \int\limits_{t_1}^{t_2}dt
 \int\limits_{V_0 }^{ } d^3x \left(\sum\limits_{f}p_f\dot f
-P_{\vh}\dot \vh -N_{0}
\left[-\frac{P_{\vh}^2}{4}+\rho(\vh)\right]\right),~~
\dot f = \frac{\partial f}{\partial t},
\eea
where $\rho(\vh)$ is the global energy density that  generates all
the above-mentioned linear equations for independent degrees of
freedom. This energy-constrained theory contains the Friedmann-like
equation for the conformal time (\ref{cst})
\be \label{time}
\eta(\vh_0,\vh_I)=
\pm\int\limits_{\vh_I }^{\vh_0 }\frac{d\vh}{\sqrt{\rho(\vh)}}
\ee
as a consequence of the energy constraint
$-{P_{\vh}^2}/{4}+\rho_F=0$
and the equation  for the dilaton momentum $P_{\vh}$
\bea \nonumber
\frac{d\vh}{d\eta}=\frac{P_{\vh}}{2}~=\pm \sqrt{\rho(\vh)}~.
\eea
The cosmic evolution of dilaton masses leads to
the redshift of energy levels of star atoms~\cite{N}
with the energy density $\rho(\vh)$ and the Hubble parameter
$H_0=\frac{\vh'}{\vh}(\eta_0)$, which gives the present-day value
of the dilaton
\bea \nonumber
 \vh(\eta_0) = \vh_0 =
\frac{\sqrt{\rho_0}}{H_0}=\Omega_0^{1/2} M_{\rm Planck}
\sqrt{\frac{3}{8\pi}},~~~~
\Omega_0 \equiv \frac{\rho_0}{{\vh_0}^2 {H_0}^2} = 1.
\eea
Therefore, the Planck scale is distinguished as a current
(present-day) value of the dilaton, rather than the fundamental
parameter that can be shifted to the beginning of the universe.
The energy-constrained theory solves also the problem of horizon by
 perturbation theory in conformal space (\ref{cst}).
If we introduce  "particles" as holomorphic field variables
\bea\nonumber
   f(t,\vec x)=\sum\limits_{k,\sigma}^{ }
\frac{C_f(\vh)\exp(ik_ix_i)}{V_0^{3/2}
\sqrt{2 \omega_f(\vh,k)}}
\left( a_{\sigma}^+(-k,t)\epsilon_{\sigma}(-k)+
a_{\sigma}(k,t)\epsilon_{\sigma}(k)\right),
\eea
where
\bea\nonumber
C_h(\vh) = \frac{\sqrt{12}}{\vh},~~~
C^{||}_{v}(\vh)=\frac{\omega_{v}}{y_{v}\vh},~~~
C_{\gamma}(\vh)=C_{s}(\vh)=C^{\bot}_{v}(\vh)=1,
\eea
then the energy density can be represented in the diagonal form
$ {\hat \rho}(\vh)=
 \sum_{\varsigma }^{ }\omega_{f}(\vh,k){\hat N}_{\varsigma}$
(where $\omega_{f}(\vh,k)=(k^2+y_f^2\vh^2)^{1/2}$ is the
one-particle energy; ${\hat N}_{\varsigma} =
\frac{1}{2}(a_{\varsigma}^{+}a_{\varsigma} +
a_{\varsigma}a_{\varsigma}^{+})$ is the number of particles;
$\varsigma$ include momenta $k_i$, species
$f=h,\gamma,v,s$, spins $\sigma$).
At the same time, the canonical differential form in the action
acquires nondiagonal terms as sources of cosmic creation of particles
\bea \nonumber
\left[\int\limits_{V_0 }d^3x \sum\limits_{f}p_f\dot f \right]_B
=\sum_{\stackrel{\scriptstyle\varsigma~ =}
{\scriptstyle (k, f,\sigma)}}
\frac{\imath}{2}( a^{+}_{\varsigma} {\dot a}_{\varsigma}
-a_{\varsigma}{\dot a}^{+}_{\varsigma} ) - \sum_{\varsigma}
\frac{\imath}{2}(a^{+}_{\varsigma}a^{+}_{\varsigma} -
a_{\varsigma}a_{\varsigma}) \dot \Delta_{\varsigma}(\vh).
\eea
The number of created particles is calculated by diagonalization
of the equations of motion by the Bogoliubov transformation
\be \label{bogtr}
b_{\varsigma}=\cosh(r_{\varsigma})
e^{ i \theta_{\varsigma}}a_{\varsigma}
 + \imath\sinh(r_{\varsigma})
 e^{-i \theta_{\varsigma}}a_{\varsigma}^{+}.
\ee
The equations for the  Bogoliubov coefficients
\bea
[\omega_{\varsigma} - \theta'_{\varsigma}] \sinh(2r_{\varsigma})&=&
\Delta'_{\varsigma}\cos(2\theta_{\varsigma})\cosh(2r_{\varsigma}),\nonumber \\
r'_{\varsigma} &=& - \Delta'_{\varsigma}\sin(2\theta_{\varsigma})\nonumber
\eea
determine the number of particles
${\it N}_{\varsigma}^{(B)}(\eta) =
{}_{sq} \langle 0|\hat N^{(B)}_{\varsigma}|0\rangle_{sq}-
{1}/{2}=\sinh^2 r_{\varsigma}(\eta)$
created during the time $\eta$ from squeezed vacuum:
$b_{\varsigma}|0\rangle_{sq} = 0$ and the evolution of the density
$\rho(\vh)=\vh'^{2} = \sum_{\varsigma} \omega_{\varsigma}(\vh)
{}_{sq}\langle 0|\hat N_{\varsigma}|0\rangle_{sq}~.$
The set of nondiagonal terms in SM
\bea \nonumber
\Delta_{h}(\vh) = \ln(\vh /\vh_{I}),
~~~\Delta^{{\bot}}_{v}(\vh) =
\frac{1}{2}\ln(\omega_{v} / \omega_{I}),~~~
\Delta^{||}_{v}(\vh) =\Delta_{h}(\vh) - \Delta^{\bot}_{v}(\vh),
\eea
where $\vh_I$ and $\omega_I$ are initial values, contains the zero-mass
singularity~\cite{sf,hp} that plays an important role in the
explanation of the longitudinal vector bosons and the origin
of the cosmic microwave background (CMB) radiation.

\section{Creation of vector bosons}
We consider the rigid state in the conformal version
${\rho}/{\rho_0}=\Omega_{\rm rigid}(z+1)^2$
with the conformal-invariant equation for the dilaton
\bea \nonumber
%(\vh^2)''=0~~~ \Rightarrow ~~~
\vh^2(\eta)=\vh^2_I[1+ 2 H_I \eta]=
\frac{\vh^2_0}{(1+z)^2}~,~~~~
\nonumber~H(z)=\frac{\vh'}{\vh}=H_0(1+z)^2~,
%~~~~~~~~~~~~~~~~~~~~~~~~~~~~~~~~~~
\eea
where $\vh_I, H_I$ are  primordial data.
At the point of coincidence of the Hubble
parameter with the mass of vector bosons
$~m_v(z)$ $\sim H(z)$, there occurs
the intensive creation of  longitudinal bosons \cite{114}.

The numerical solutions of the Bogoliubov equations
 for the time dependence of the vector boson distribution
functions ${\it N}_v^{||}(k,\eta)$ and ${\it N}_v^{\bot}(k,\eta)$ are given
in Fig. 1 (left panels) for the momentum $k=1.25 H_I$.
We  can see that the longitudinal function
is noticeably greater than the transversal one.
The momentum dependence of these functions at the beginning
of the universe is given on the right panels of Fig.1.
The upper panel shows us the intensive cosmological creation of
the longitudinal bosons in comparison with
the transversal ones.
This fact is in agreement with the mass singularity of
the longitudinal vector bosons discussed in~\cite{sf,hp}.
One of the features of this intensive creation is a high momentum
tail of the momentum distribution of longitudinal bosons which leads
to a divergence of  the density of created particles defined
as~\cite{par}
\be\label{nb}
n_{v}(\eta)=\frac{1}{2\pi^2}
\int\limits_{0 }^{\infty } dk k^2
\left[ {\it N}_v^{||}(k,\eta) + 2{\it N}_v^{\bot}(k,\eta)\right]~.
\ee
The divergence is a defect of our approximation
where we neglected all interactions of vector bosons that form
the collision integral in the kinetic equation for the distribution
functions.
In order to obtain a finite result for the density, we suggest
multiplying the primordial distributions
${\it N}_v^{||}(k,\eta_L)$ and ${\it N}_v^{\bot}(k,\eta_L)$ with
the Bose - Einstein distribution $(k_B=1)$
\be \label{bose}
{\it F}_v (k,\eta)=
\left\{\exp\left(\frac{\omega_v(\tau)-m_v(\eta)}{T}\right)
-1\right\}^{-1}\!\!\!\!,
\ee
where $T$ is considered as a regularization parameter.

Our calculation of this density presented in Fig.1 signals that the
density~(\ref{nb}) is established very quickly in comparison with
the lifetime of bosons, and in the equilibrium
there is a weak dependence of the density on the time (or z-factor).
This means that the initial Hubble parameter $H_I$  almost
coincides with the Hubble parameter at the point of saturation $H_s$.

\begin{figure}
 \includegraphics[width=0.98\textwidth,clip]{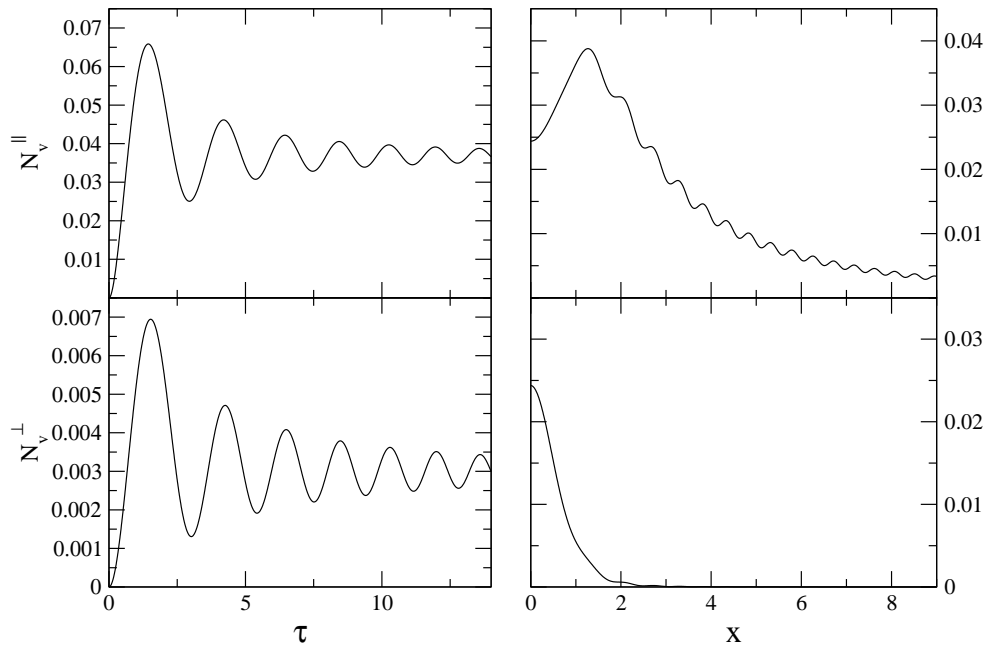}
 \caption{Time dependence  for the dimensionless momentum
 $x=k/H_I=1.25$ (left panels) and momentum dependence, at the
 dimensionless lifetime $\tau=(2 \eta H_I)=14$, (right panels)
 of the transverse (lower panels) and longitudinal (upper panels)
 components of the vector-boson distribution function.}
\end{figure}

For example, we have calculated the values of integrals (\ref{nb})
for the regularization parameter
$T= m_v(z_I)= H_I$.
The result of the calculation is
\be \label{nc}
\frac{n_{v}}{T^3} = \frac{1}{2\pi^2}
\left\{ [1.877]^{||}+2 [0.277]^{\bot}=2.432   \right\}~.
\ee
The final product of the decay of the primordial bosons includes
photons. If one photon comes from the annihilation of the products
of the decay of $W^{\pm}$ bosons and another from $Z$ bosons,
we can expect the density of photons in the conformal cosmology with
the constant temperature and the static universe~\cite{039} to be
\be \label{1nce}
\frac{n_{\gamma}}{T^3}=\frac{1}{\pi^2}
\left\{ 2.432  \right\}.
\ee
Comparing this value with the present-day density of
the cosmic microwave radiation
\be \label{1nc}
\frac{n^{\rm obs}_{\gamma}}{T_{\rm CMB}^3}=\frac{1}{\pi^2}
\left\{ 2 \zeta (3)= 2.402 \right\}
\ee
we can estimate the regularization parameter $T$.
One can see that this parameter is of the order the
temperature of the cosmic microwave background
$T=T_{\rm CMB}= 2.73~{\rm K}$.
We can speak about thermal equilibrium for the primordial bo\-sons
with the temperature $T_{\rm eq}$, if the  inverse relaxation
time~\cite{ber}
$\tau_{\rm rel}^{-1}(z_I) = {\sigma_{\rm scat }n_{v}(T_{\rm eq})}$,
where the scattering cross-section of bosons in the considered region
proportional to the inverse of their squared mass
$\sigma_{\rm scat}={\gamma_{\rm scat}}/{m^2_v(z)}$
is greater than the primordial Hubble parameter $H_I$.
This means that the thermal equilibrium will be maintained, if
\be \label{rel01}
\gamma_{\rm scat}n_v(T_{\rm eq }) =
\frac{2.4~\gamma_{\rm scat}}{2\pi^2}T_{\rm eq}^3
> H(z_I)m^2_v(z_I).
\ee
The right-hand side of this formula is an integral of motion for
the evolution of the universe in the rigid state. The estimation
of this integral from the present values of the Hubble parameter
and boson mass gives the value
\be \label{tcmbr1}
 \left[m_W^{2}(z_I)H(z_I)\right]^{1/3}
 =\left[m_W^{2}(0)H_0\right]^{1/3} = 2.76~{\rm K}.
\ee
justified, when for the coefficient $\gamma_{\rm scat}>8.5$ holds.
 
Thus, we conclude that the assumption of a quickly established
thermal equilibrium in the primordial vector boson system may be
justified, since $T\sim T_{\rm eq }$. The temperature of the photon
background emerging after annihilation and decay processes
of $W^{\pm}$ and $Z$ bosons is invariant in the conformal cosmology
and the simple estimate performed above gives a value
surprisingly close to that of the observed CMB radiation.

\section{Nonconservation of fermion quantum numbers}
The interaction of  primordial W and Z bosons with left-handed fermions
leads to nonconservation of the fermion quantum numbers. It is
well known that the gauge-invariant current of each doublet is
conserved at the classical level, but not the quantum one \cite{ufn}.
At the quantum level, we have the anomalous current
$j_L^{(i)}=\bar\psi^{(i)}_L\gamma_\mu\psi^{(i)}_L$,
\bea \nonumber
\partial_\mu j_L^{(i)}=-\frac{{\rm
Tr}F_{\mu\nu}{}^*\!{F_{\mu\nu}}}{16\pi^2}, ~~
F_{\mu\nu}=-\frac{\imath g\tau_a}{2}{F_{\mu\nu}^{a}}^{\rm(p.t.)}, ~~
{F_{\mu\nu}^{a}}^{\rm(p.t.)}
=\partial_\mu A_\nu^a-\partial_\nu A_\mu^a.\eea

If we take the integral over four-dimensional conformal
space-time confined between three-dimensional hyperplanes
$\eta=0$ and $\eta=\eta_L$, we find that the number of left fermions
$N(\eta_L)$ is equal to the difference of Chern-Simons numbers
 $\Delta N_{CS}(\eta_L)=N_{CS}(\eta_L)-N_{CS}(0)$
\bea \nonumber N(\eta_L)=
 -\int_0^{\eta_L} d\eta \int \frac{d^3 x}{16\pi^2} \;
 {}_{\rm sq}\langle 0|{\rm Tr}F_{\mu\nu}
 {}^*\!{F_{\mu\nu}}|0\rangle{}_{\rm sq} =\Delta N_{CS}(\eta_L)~,
\eea
 where $\tau_L=\eta_L/\eta_I$ is the lifetime of bosons.
 To estimate this time $\eta_L$, we use the lifetime of
W-bosons in the Standard Model at this moment
\bea \nonumber
\eta_L+\eta_I= \frac{\sin^2 \theta_W}{m_W(z_L)\alpha_{QED}}~,
\eea
where $\theta_W$ is the Weinberg angle,  $\alpha_{QED}=1/137$,
and $z_L$ is the z-factor at this time. Using
the equation of the rigid state \cite{114} and the equality $\eta_I
m_W(z_L)=(z_I+1)/(2[z_L+1])$, we rewrite the previous
equation in terms of the z-factor \be \label{life}
\tau_L+1=\frac{(z_I+1)^2}{(z_L+1)^2}= \frac{(z_L+1)}{(z_I+1)}
\frac{2\sin^2\theta_W}{\alpha_{QED}}~.
\ee
The solution of this equation is
$ \tau_L+1=({2\sin^2\theta_W}/{\alpha_{QED}})^{2/3}\simeq 15$,
and for the lifetime of
created bosons we have $ \tau_L={\eta_L}/{\eta_I}\simeq 14$.

During their lifetime, transverse vector bosons are evolving so
that the Chern-Simons functional is changed
\be 
\label{nw} 
\Delta N_w =
\frac{4{\alpha}_{QED}}{\sin^{2}\theta_{W}}\int_{0}^{\eta^{\rm W}_{\rm l}} 
d\eta \int \frac{d^3 x}{4\pi}~~{}_{\rm sq}
\langle 0|E^{W}_{i}B^{W}_{i}|0\rangle{}_{\rm sq}~,
\ee
where $E_i$ and $B_i$ are the electric and magnetic fields strengths.
The squeezed vacuum and Bogoliubov transformations (\ref{bogtr}) give
a nonzero value for these quantities
\be 
\label{eb} 
\int  \frac{d^3 x}{4\pi}~{}_{\rm sq}
\langle 0|E^{v}_{i}B^{v}_{i}|0\rangle{}_{\rm sq}
= -\frac{V_0}{2} \int\limits_{0 }^{\infty }dk |k|^3
\cos(2\theta_\zeta)\sinh(2r_\zeta),
\ee
where $\theta_\zeta$ and $r_\zeta$ are given by the equation for
transverse bosons.  Using  the relation
\be 
\label{nz} 
\Delta N_Z =
\frac{{\alpha}_{QED}}{\sin^{2}\theta_{W}\cos^{2}\theta_{W}}
\int_{0}^{\eta^{\rm Z}_{\rm l}} d\eta \int
\frac{d^3 x}{4\pi}~~{}_{\rm sq}
\langle 0|E^{Z}_{i}B^{Z}_{i}|0\rangle{}_{\rm sq}, 
\ee
we find the Chern-Simons functional for Z bosons in a similar way.
 We estimate
\bea 
\nonumber
\frac{(3\Delta B)}{V_0T^3}=
\frac{(\Delta N_W+\Delta N_Z)}{V_0T^3}=
 \frac{{\alpha}_{QED}}{\sin^{2}\theta_{W}}
  \left(4\times 1.44+\frac{2.41}{\cos^{2}\theta_{W}}\right)
\eea
for  the lifetime of bosons $\tau_L^W\approx 15$,
$\tau_L^Z\approx 30$, using the T-regularization~(\ref{bose}).

The baryon asymmetry, which appeared as a consequence of three Sa\-kha\-rov
conditions ( $CP_{\rm SM}$, $H_0\not= 0$,
 $\Delta L = 3 \Delta B = \Delta n_w + \Delta n_z$),
is equal to ${\Delta B}/{n_\gamma}=X_{CP}/3$
where $X_{CP}$ is a factor determined by a superweak interaction
of $d$ and $s$-quarks $(d+s~\rightarrow ~s+d)$
with CP-violation experimentally observed in decays of
$K$ mesons~\cite{o}.

It is worth emphasizing that in the considered model of the
conformal cosmology, the temperature is a constant. In conformal
cosmology, we have the mass history 
\be 
\label{mz} 
m_{\rm era}{(z_{\rm era})}=
\frac{m_{\rm era}(0)}{(1+z_{\rm era})}=T_{\rm eq},
\ee
with the constant temperature $T_{\rm eq}= 2.73 ~ {\rm K} = 2.35
 \times 10^{-13}$ GeV
where $m_{\rm era}(0) $ is characteristic energy (mass) of the
era of the universe evolution, which begins at the redshift $z_{\rm era}$.

Eq. (\ref{mz}) has the important consequence that all
physical processes, which concern the chemical composition of the
universe and depend basically on the Boltzmann factors with the
argument $(m/T)$, cannot distinguish between the conformal
cosmology and the standard cosmology due to the relations
\be
\nonumber
\frac{m(z)}{T(0)}=\frac{m(0)}{(1+z)T(0)}=\frac{m(0)}{T(z)}~. 
\ee
This formula makes it transparent that in this order of approximation
a $z$-history of masses with invariant temperatures in the rigid
state of conformal cosmology is equivalent to a $z$-history of
temperatures with invariant masses in the radiation stage of the
standard cosmology. We expect, therefore, that the conformal
cosmology allows us to keep the scenarios developed in the
standard cosmology in the radiation stage for, e.g. the
neutron-proton ratio and primordial element abundances.

\section{Conclusion}
We have considered the unified theory of all interactions in the
space-time with the Weyl relative standard of measurements.
The conformal symmetry, reparametrization - invariant
perturbation theory, and the mass-singularity of longitudinal
components of vector bosons lead to the effect of intensive
creation of these bosons with the temperature of an order of
$(m_W^2H_0)^{1/3} \sim  2.73 ~ {\rm K}$.
Instead of the z-dependence of the temperature in
an {\it expanding universe} with  constant masses in the standard
cosmology, in conformal cosmology, we have the z-history of masses
in a {\it non-expanding universe} with an almost constant temperature of
the photon background (with the same argument of the Boltzmann
factors).
Recall that the density resemble physical
properties of the cosmic microwave background radiation.
The primordial boson "radiation" created during a conformal
time interval of $2 \times 10^{-12} {\rm sec}$ violates
the baryon number. The subsequent
annihilation and decay of primordial bosons
form all the matter in the universe in the rigid state.
At the present-day stage, the evolution of the universe in the
rigid state in the conformal cosmology does not contradict recent
observational data on Supernova \cite{ps} and confirms the results
on the abundance of chemical elements obtained in the Hot Scenario
at the radiation stage~\cite{114}.

\section*{  Acknowledgments}
The authors are grateful to  E.A.~Kuraev, S.I.~Vinitsky, and
participants of the Seminar of the Russian Gravitation Society for
fruitful discussions. DP and AG thank RFBR (grant 00-02-81023 Bel
2000$\_a$) for support.

\end{document}